\documentclass[aps,preprint]{revtex4}
\usepackage{graphicx}
\begin{document}
\title{\Large \bf Charged Rotating Black Holes on a 3-Brane}
\author{\large A. N. Aliev}
\affiliation{ Feza G\"ursey Institute, P.K. 6  \c Cengelk\" oy,
81220 Istanbul, Turkey}
\author{\large A. E. G\"umr\"uk\c{c}\"uo\~glu}
\affiliation{ITU, Faculty of Sciences and
Letters, Department of Physics, 34469 Maslak, Istanbul, Turkey}
\date{\today}
\begin{abstract}
\noindent We study exact stationary and axisymmetric solutions
describing charged rotating black holes localized on a 3-brane in
the Randall-Sundrum braneworld. The charges of the black holes are
considered to be of two types, the first being an induced {\it
tidal} charge that appears as an imprint of nonlocal gravitational
effects from the bulk space and the second is a usual electric
charge arising due to a Maxwell field trapped on the brane. We
assume a special ansatz for the metric on the brane  taking it to be
of the Kerr-Schild form and show that the Kerr-Newman solution of
ordinary general relativity in which the electric charge is
superceded by a tidal charge satisfies a closed system of the
effective gravitational field equations on the brane. It turns out
that the negative tidal charge may provide a mechanism for spinning
up the black hole so that its rotation parameter exceeds its mass.
This is not allowed in the framework of general relativity. We also
find a new solution that represents a rotating black hole on the
brane carrying both charges. We show that for a rapid enough
rotation the combined influence of the rotational dynamics and the
local bulk effects of the "squared" energy momentum tensor on the
brane distort the horizon structure of the black hole in such a way
that it can be thought of as composed of non-uniformly rotating null
circles with growing radii from the equatorial plane to the poles.
We finally study the geodesic motion of test particles in the
equatorial plane of a rotating black hole with tidal charge. We show
that the effects of negative tidal charge tend to increase the
horizon radius, as well as the radii of the limiting photon orbit,
the innermost bound and the innermost stable circular orbits for
both direct and retrograde motions of the particles.

\end{abstract}
\maketitle
\section{Introduction}

Black holes have been among the most intriguing objects of study for
many years in both ordinary general relativity and higher
dimensional gravity theories. Recently, after the advent of
braneworld gravity theories, the interest in black holes gained new
impetus  due to striking phenomenological consequences of the
braneworld models \cite{ADD}-\cite{RS2} (see also Refs.\cite{rsh}
for simple field-theoretical models). A comprehensive description of
the different aspects of the subject can be found in recent reviews
\cite{roy}-\cite{rubakov}.

The basic ingredient of the braneworld models is that our physical
Universe is a $3$-brane with all matter fields localized on it,
while gravity is free to propagate in all extra spatial dimensions.
The size of the extra dimensions may be much larger \cite{ADD} than
the conventional Planck length scale ($ L \gg l_{P} \simeq 10^{-33}
cm $). The extra dimension may even have an infinite size, like in
the braneworld model of Randall and Sundrum (RS) consisting of a
single extra spatial dimension \cite{RS2}. One of the dramatic
consequences of the braneworld models is that they may provide one
with an elegant approach to the solution of the hiearchy problem
between the electroweak scale and the fundamental scale of quantum
gravity. Namely, due to the large size of extra dimensions the scale
of quantum gravity becomes the same order as the electroweak
interaction scale (of the order of a few TeVs). This, in turn, opens
up the possibility of TeV-size black hole production at high energy
collision processes in cosmic ray and at future colliders
\cite{emp1}-\cite{dl}.

The braneworld model with an infinite extra dimension \cite{RS2}
realizes the localizaton of a 5-graviton zero mode on a 3-brane
along with rapid damping at large distances of massive Kaluza-Klein
(KK) modes. Thus, the conventional potential of Newtonian gravity is
recovered on the 3-brane with high enough accuracy. Moreover, it has
been shown that the Randall-Sundrum braneworld model also supports
the properties of four-dimensional Einstein gravity in low energy
limit \cite{gt}- \cite{sms1}. In light of all this, it is natural to
assume the formation of black holes in the braneworld due to
gravitational collapse of matter trapped on the brane. Such black
holes would be mainly localized on the brane, however some portion
of their event horizon would have a finite extension into the extra
dimension as well. In other words, they will in general manifest
themselves as higher dimensional objects.

Several strategies have been discussed in the literature to describe
the braneworld black holes. First of all, it is clear that if the
radius of the horizon of a black hole on the brane is much smaller
than the size scale of the extra dimensions $(\,r_{+} \ll L \,)$,
the black hole will "feel" higher dimensional spacetime as a whole.
In this case the braneworld black holes, to a good enough
approximation, can be described by the classical solutions of higher
dimensional vacuum Einstein equations found a long time ago
\cite{tang}-\cite{mp}. (For the recent study of different physical
properties of these solutions  see Refs.\cite{fs1}-\cite{saka}).
Interesting examples of numerical solutions describing small black
holes localized on a 3-brane were also given in \cite{ktn}. In the
opposite case, when the horizon radius is much greater compared to
the length scale of the extra dimensions ($\,r_{+} \gg L \,)$, the
black hole becomes effectively four-dimensional with a finite
extension along the extra dimensions. The exact description of such
black holes is much more difficult and it still needs to be found.
However, such a description is possible for black holes in a
three-dimensional Randall-Sundrum braneworld model
\cite{ehm1}-\cite{ehm2}.

The authors of paper \cite{chr} have proposed a simple black hole
solution in the RS braneworld model with infinite size extra
dimension. They suggested that the endpoint of gravitational
collapse of non-rotating matter trapped on the brane could still be
well described by a usual Schwarzschild metric, however it would
look like a {\it black string} solution from the point of view of an
observer in the bulk. Unfortunately, the black string solution
turned out to be plagued by curvature singularities at infinite
extension along the extra dimension (AdS horizon of the RS
braneworld). Though there has been some speculations \cite{chr} that
the black string solution may evolve to a localized {\it black
cigar} soluton due to its classical instability near the AdS horizon
\cite{glaf}, the question of singularites was not resolved
ultimately \cite{hm}. The similar case of a rotating black string
solution was considered in \cite{modgil}, while the solutions with
dilaton field were given in \cite{nojiri}.

Another strategy of finding an exact localized black hole solution
in the braneworld was initiated in \cite{dmpr}. The authors
specified the metric form induced on the 3-brane assuming a
spherically symmetric ansatz for it and then solved the effective
gravitational field equations on the brane. The solution turned out
to be a Reissner-Nordstrom type static black hole possessing a
"tidal" charge instead of a usual "electric" charge. The tidal
charge has five dimensional origin and can be thought of as an
imprint of the gravitational effects from the bulk space.
Subsequently, this approach has been repeatedly used by several
authors to describe different, localized on a 3-brane, static black
hole solutions \cite{gm}-\cite{rw} (see also \cite{roy} for a
review). Some observational signatures of the braneworld black
holes, like gravitational lensing were also studied in \cite{whis}.

However, the exact bulk metrics for these types of solutions have
not been constructed yet. The numerical integration of the field
equations into the bulk, using the exact localized static black hole
solutions as "initial  data", runs into difficulties \cite{chrss}.

In this paper we study exact stationary and axisymmetric solutions
describing charged rotating black holes localized on a 3-brane in
the Randall-Sundrum braneworld. We shall consider two types of the
charged solutions. The first one is the Kerr-Newman type metric for
a rotating black hole on the brane, in which instead of a usual
Maxwell charge contribution, a {\it five} dimensional correction
term from the bulk space appears. The latter arises due to the
gravitational "coupling"  between the brane and the bulk, which on
the brane is felt through the "electric" part of the bulk Weyl
tensor. Therefore, as in the case of \cite{dmpr}, this type of exact
solution can be re-interpreted as representing a rotating braneworld
black hole with a {\it tidal} charge. The second solution represents
a rotating braneworld black hole which in addition to the tidal
charge, also carries an electric charge arising due to a Maxwell
field on the brane.

It is interesting to note that the solution with both types of
charges does not admit the Kerr-Newman form in the usual sense.
Instead, in addition to the off-diagonal component of the metric
with indices $\,t\,$ and $\,\phi\,,$  we also have the off-diagonal
components of the metric with indices $\,(r \phi)\,$ and $\,(r t)\,$
that arise due to the combined influence of the local bulk effects
on the brane and the \textit{dragging effect} of the rotating black
hole. Furthermore, the combined influence of rotational dynamics of
the black hole and the local bulk effects of the "squared" energy
momentum tensor on the brane distort the event horizon of the black
hole in such a way that it becomes a stack of non-uniformly rotating
null circles extending from the equatorial plane to the poles and
having different radii at different, fixed $\,\theta\,$. However,
for small enough values of the rotation parameter (i. e. in linear
approximation) the horizon rotates with a uniform angular velocity
and the metric goes over into the usual Kerr-Newman form for a
charged and slowly rotating black hole on the brane. On the other
hand, in the absence of rotation the metric reduces to the
Reissner-Nordstrom type solution with correction terms of local and
nonlocal origin \cite{chrss} .

The paper is organized as follows. In Sec.II we begin with a brief
description of the effective gravitational field equations induced
on a 3-brane in the five-dimensional bulk space. In Sec.III we
specialize to the case of the Randall-Sundrum braneworld scenario
and propose a special metric ansatz  for a non-charged rotating
black hole on the brane. Namely, we assume that the metric on the
brane can be taken in the Kerr-Schild form and solve the Hamiltonian
constraint equation for this metric form. Then we calculate the
"electric" part of the bulk Weyl tensor which provides one with the
above choice for the metric ansatz. The solution for a rotating
black hole on the brane and carrying both tidal and electric
charges, as well as the analytical and numerical study of its
properties are given in Sec.IV. Finally, in Sec.V, we study the
geodesic motion of test particles in the equatorial plane of a
rotating black hole with tidal charge. We obtain the basic equations
governing the radius of the limiting photon, the radii of the
marginally bound and the marginally stable circular orbits and apply
the numerical analysis to explore the effect of a negative tidal
charge on these orbits.

\section{Gravitational Field Equations on the brane}

The effective gravitational field equations on a 3-brane in a
five-dimensional bulk spacetime possessing  a $Z_{2}$ mirror
symmetry  with respect to the brane were derived by Shiromizu, Maeda
and Sasaki \cite{sms1} using the Gauss-Codazzi projective approach
with the subsequent specialization to the Gaussian normal
coordinates. Recently, in \cite{aemir} it has been shown that the
field equations on the brane remain unchanged even in more general
coordinate setting which allows for acceleration of the normals to
the brane surface through the lapse function and the shift vector in
the spirit of Arnowitt, Deser and Misner (ADM) formalism \cite{adm}
of four-dimensional general relativity . However, the evolution
equations into the bulk are significantly changed in the ADM
setting. We start with a brief description of the gravitational
field equations on the brane in the spirit of the ADM approach.

We consider a five-dimensional bulk spacetime with metric $ g_{AB} $
and suppose that the spacetime includes a $3$-brane which in turn is
endowed with the metric $h_{AB}$. Assuming that the bulk spacetime
is covered by coordinates $ x^{A}$ with $A=0,1,2,3,5$ we write down
the Einstein field equations in the form
\begin{eqnarray}
^{(5)}G_{AB}&=& \, ^{(5)}R_{AB}- \frac{1}{2}\,g_{AB}\,^{(5)}R \,=\,
- \Lambda_{5} \, g_{AB}+
\kappa_5^2\left(^{(5)}T_{AB}+\sqrt{\frac{h}{g}}\,\,\tau_{AB}\,
\delta(Z)\right)\,, \label{5dfeq}
\end{eqnarray}
where $\,\kappa_{5}^2=8\pi G_5\,$, with $\,G_5\,$ being the
gravitational constant, $\Lambda_{5}$ is the bulk cosmological
constant, while  $^{(5)}T_{AB}$ is the energy-momentum tensor in the
bulk, $\,\tau_{AB}\,$ is the energy-momentum tensor on the brane and
the quantities $\,g\,$ and $\,h\,$ are the metric determinants of
$g_{AB}$ and $h_{AB}$, respectively.

The effective field equations on the brane are obtained by using a
$(4+1)$ slicing of the five-dimensional bulk spacetime. We introduce
an arbitrary scalar function
\begin{equation}
Z= Z(x^A) \,, \label{scalarf}
\end{equation}
such that $Z$= const describes a given surface in the family of
non-intersecting timelike hypersurfaces $\Sigma_Z$, filling the
bulk. We assume that the $3$-brane is located at the hypersurface
$Z=0$\,. Clearly, one can also introduce the unit spacelike normal
to the brane surface
\begin{eqnarray}
n_A &=& N\,\partial_A Z =\left(0,0,0,0,N \right)\
 \label{vnormal}
\end{eqnarray}
satisfying the normalization condition
\begin{equation}
g_{AB}\,n^A\,n^B = 1\,, \label{normalization}
\end{equation}
where  the scalar function $N=\left| g^{AB}\,\partial_A
Z\,\partial_B Z \right|^{-1/2}$ is  called the lapse function. Next,
we appeal to the parametric equation of the brane surface $ x^A=x^A
(y^i) $. This implies the existence of a local frame given by the
set of four vectors
\begin{equation}
e^A_i = \frac{\partial x^A}{\partial y^i}~~~, \label{tvector}
\end{equation}
which are tangent to the brane. Here the coordinates $y^{i}$ with
$i=0,1,2,3$ are intrinsic to the brane.
 It is clear that
\begin{equation}
\,n_A\,e^A_i = 0 \,. \label{orthogonal}
\end{equation}
With this local frame the infinitesimal displacements on the brane
are determined by the induced metric
\begin{equation}
h_{ij} = g_{AB} \,e^A_i \,e^B_j\,\,, \label{indmetric}
\end{equation}
From equations (\ref{orthogonal}) and (\ref{indmetric}) it follows
that the bulk spacetime metric can be represented as
\begin{equation}
g_{AB} = n_A n_B +h_{ij} e^i_A e^j_B \,\,, \label{5metric1}
\end{equation}
We also need to introduce a spacelike vector $Z^A$, which formally
can be thought of as an "evolution vector" into the fifth dimension.
Having chosen the parameter along the orbits of this vector as Z, we
have the relation
\begin{equation}
 Z^A\,\partial_A Z =1\,, \label{evector1}
\end{equation}
which means that the vector $Z^A$ is tangent to a congruence of
curves intersecting the succesive hypersurfaces in the slicing of
spacetime. Clearly, in general case the evolution vector $Z^A$ can
be decomposed into its normal and tangential parts
\begin{equation}
Z^A = N\,n^A +N^i\,e^A_i \,\,, \label{evector2}
\end{equation}
where the four vector $N^i$ is known as the shift vector (see
Ref.\cite{eric}). With this construction one can always define an
alternative coordinate system $\,(y^i, y^5)\equiv (y^i, Z)\,$ on the
bulk space, and then decompose the spacetime metric in these new
coordinates as
\begin{eqnarray}
ds^2&=&g_{AB} dx^{A}dx^{B} \nonumber \\
&=& h_{ij}\, dy^i\,dy^j + 2N_i \,dy^i \,dZ + \left(N^2 +N_i N^i
\right)\,dZ^2 \,. \label{demetric}
\end{eqnarray}

The bending of the brane surface in the bulk is described by the
symmetric extrinsic curvature tensor $\,K_{AB}\,$ which can be
introduced through the relations
\begin{eqnarray}
\nabla_A\,n_B&=&K_{AB}+n_A a_B\,,\,\,\,\, \,K_{AB} n^A=0 \,,
\label{acceleration1}
\end{eqnarray}
where $\,\nabla\,$ is the covariant derivative operator associated
with the bulk metric $\,g_{AB}\,$ and the 5-acceleration of the
normals is given by
\begin{equation}
a_A = n^B \,\nabla_B\,n_A\,\,.\label{acceleration}
\end{equation}
It is also useful to present the extrinsic curvature tensor in its
projected form
\begin{eqnarray}
K_{ij}&=& \nabla_{(B}n_{A)}\, e^A_i e^B_j\,\ =\frac{1}{2N} \left(
\partial_{5} h_{ij}- D_i N_j-D_j N_i\right)\,, \label{excurv1}
\end{eqnarray}
where $\,\partial_5= \partial/\partial{Z}\,\,$ and $D$ is the
covariant derivative operator defined with respect to the induced
metric $\,h_{ij}\,$ on the brane.

The effective Einstein field equations on the brane arise in three
steps (for details see Ref.\cite{aemir}); i) the derivation of the
reduction formulas between the components of the five-dimensional
Einstein tensor in equation (\ref{5dfeq}) and the quantities
characterizing the intrinsic and extrinsic curvature properties of
the brane surface in accord with the decomposition (\ref{demetric})
of the bulk metric, ii) the relation of the $\delta$-function
singularity on the right-hand-side of equation (\ref{5dfeq}) to the
jump in the extrinsic curvature of the brane on its evolving into
the fifth dimension, iii) the expression of the corresponding
evolutionary terms in the resulting equation in terms of
four-dimensional quantities on the brane, more precisely in terms of
their limiting values on the brane. Having done all this and
assuming that the braneworld energy-momentum tensor has the
particular form $\,\tau_{ij} = - \lambda h_{ij} +T_{ij}\,,$ for
which the Israel's \cite{Israel} junction condition on the $Z_2$
symmetric brane results in the relation
\begin{equation}
K_{ij}=-\frac{1}{2}\,\kappa_5^2\,\left[ T_{ij}-
\frac{1}{3}\,h_{ij}\,\left(T-\lambda \right)\,\right]\,,
\label{brextcur}
\end{equation}
where $\,\lambda\,$ is the brane tension, we arrive at the
gravitational field equations on the brane
\begin{equation}
G_{ij} = -\Lambda h_{ij} + \kappa_4^2 \, T_{ij} + \kappa_5^4
\,S_{ij} - W_{ij} - 3 \kappa_5^2\,U_{ij}\,\,. \label{breq1}
\end{equation}
Here we have introduced the traceless tensor
\begin{equation}
 W_{ij}=A_{ij}-\frac{1}{4}\,h_{ij}\, A
\label{wtrl}\,\,,
\end{equation}
which is constructed from the "electric" part of the bulk Riemann
tensor
\begin{equation}
A_{ij}={}^{(5)} R_{ABCD}\,n^A \,n^C e^B_i e^D_j \,, \label{defab}
\end{equation}
and its trace with respect to the metric $\,h_{ij}\,$. We also have
\begin{eqnarray}
\Lambda &=& \frac{1}{2} \left(\Lambda_5 + \frac{1}{6}\kappa_5^4
\,\lambda^2 - \kappa_5^2\, P\right)\,,
\label{cosmcons}\\
\kappa_4^2 &=& \frac{1}{6}\,\kappa_5^4\,\lambda \,\,,
\label{gravconst4}\\
S_{ij} &=& - \frac{1}{4} \left[ \left(T^m_i T_{mj} - \frac{1}{3}\,T
T_{ij}\right) - \frac{1}{2}\,h_{ij} \left(T_{lm}T^{lm} -
\frac{1}{3}\,T^2\right) \right]\,\,. \label{quademt2}
\end{eqnarray}
From the above equations we see that nonlocal gravitational effects
of the bulk space are transmitted to the brane through the quantity
$\,W_{ij}\,$, whereas the local bulk effects  on the brane are felt
through the "squared energy-momentum" tensor $\,S_{ij}\,$.
Furthermore, the influence of the bulk space energy and momentum on
the brane is described by both the normal compressive "pressure"
term $\,P=\ ^{(5)}T_{AB}\,n^A n^B\,$ and the traceless in-brane part
of the bulk energy-momentum tensor
\begin{equation}
U_{ij}= - \frac{1}{3}\,\left(^{(5)}T_{ij} - \frac{1}{4}\,h_{ij}
h^{lm}\,^{(5)}T_{lm}\right)\,\,.
\label{utrl}
\end{equation}

In addition to the above equations, the $(4+1)$ decomposition of the
five-dimensional field equations (\ref{5dfeq}) results also in the
Hamiltonian constraint equation
\begin{equation}
\frac{1}{2}\,\left(R-K^2 + K_{lm}K^{lm}\right) =\,\Lambda_{5}
-\kappa_5^2~P \,\,, \label{constr1}
\end{equation}
and the momentum constraint equation
\begin{equation}
D_m K^m_i-D_i K = \kappa_5^2~ J_i\,\,, \label{constr2}
\end{equation}
where $\,K= K _{ij} h ^{ij}\,$ and the $4$-vector
$J_{i}=^{(5)}T_{iB}\,n^B\,$ describes the energy-momentum flux onto,
or from the brane. In the case of empty bulk the above equations
recover those obtained in \cite{sms1}.

It is important to note that the consistent solutions subject to the
field equations (\ref{breq1}) on the brane require the knowledge of
the nonlocal gravitational and energy-momentum terms from the bulk
space. Thus, in general, the field equations on the brane  are
\textit{not closed} and one needs to solve the evolution equations
into the bulk for the projected bulk curvature and energy-momentum
tensors \cite{aemir}. The interesting discussions of the combined
effects of the brane-bulk dynamics on cosmological evolution on the
brane can be found in recent papers \cite{apos}-\cite{masato}.
However, in particular cases one can still make the system of
equations on the brane \textit{closed} assuming a special ansatz for
the induced metric on the brane. In what follows we shall adopt this
strategy.

\section{Rotating Black Hole with tidal charge}

We shall now construct a stationary and axisymmetric metric
describing a rotating black hole localized on a 3-brane in the
Randall-Sundrum braneworld scenario. For this purpose, we shall make
a particular assumption about the stationary and axisymmetric metric
on the brane, taking it to be of  the Kerr-Schild form. This type of
strategy was first adopted in \cite{dmpr}, where to describe a
localized static black hole configuration on the brane, the authors
assumed a spherically symmetric ansatz for the induced metric and
then solved the corresponding effective gravitational field
equations on the brane. They found a solution that turned out to be
of the form of a Reissner-Nordstrom type metric in ordinary general
relativity, however involving a "tidal" charge instead of a usual
"electric" charge. The tidal charge  has been naturally interpreted
as an imprint of the gravitational effects from the bulk space.

The Kerr-Schild ansatz \cite{ks} for the induced metric on the
brane, loosely speaking, implies that the exact metric for a
rotating black hole  on the brane can be expressed in the form of
its linear approximation around the flat metric. Thus, for the
interval of spacetime  we have
\begin{equation}
ds^2= \left(ds^2 \right)_{flat} +H\, (l_{i} dx^{i}) ^2 \, ,
\label{kerrschild1}
\end{equation}
where  $\,H\,$ is an arbitrary scalar function and $\,l_{i}\,$ is a
null, geodesic vector field in both the flat and full metrics.
Alternatively, introducing the coordinates
$\,y^i=\{u,r,\theta,\varphi\}\,$  one can go over into the following
form of the metric
\begin{eqnarray}
ds^2 &=&h_{ij} \,dy^i dy^j = \left[ - \left( du + dr \right) ^2 +
dr^2 + \Sigma \ d \theta^2 + \left( r^2 + a^2 \right) \sin^2 \theta
\,d \varphi^2 + 2\, a  \sin^2\theta \,dr d \varphi \right]
\,\nonumber\\ &&+ \,H \left(du - a \sin^2\theta \,d \varphi
\right)^2\,, \label{ansatz}
\end{eqnarray}
where the expression in square brackets corresponds to the flat
spacetime , the scalar $\,H\,$ is a function only of the coordinates
$\,r\,$ and $\,\theta\, $, the parameter $a$ is related to the
angular momentum of the black hole and the quantity $\,\Sigma\,$ is
given by
$$\Sigma = r ^2 + a ^2 \cos ^2 \theta \,.$$
Henceforth we shall suppose that the bulk space is empty and there
are no matter fields on the brane ($\, T_{ij}=0\,$). In this case
the effective gravitational field equations (\ref{breq1}) can be
reduced to the form
\begin{equation}
R_{ij}=-E_{ij}\,,
\label{breq2}
\end{equation}
where we have taken into account that for empty bulk space, the
tensor $\,W_{ij}\,$ defined in (\ref{wtrl}) is replaced by  the
traceless "electric part" of the five-dimensional Weyl tensor
\cite{sms1}
\begin{equation}
E_{ij}={}^{(5)}C_{ABCD}\,n^A n^C e^B_i e^D_j\,\,, \label{elweyl}
\end{equation}
and also the fact that in the RS model the cosmological constant
(\ref{cosmcons}) on the brane vanishes due to the fine-tuning
condition. From equations (\ref{cosmcons}) and (\ref{gravconst4}) it
follows that
\begin{eqnarray}
\Lambda_5 = - \frac{6}{\ell^2}\,\;,\;\;\;\; G_4=
\frac{G_5}{\ell}\;\;, \label{tuning}
\end{eqnarray}
where $\,\ell=6(\lambda \kappa_5^2)^{-1} \,$ is the curvature radius
of AdS . (In what follows we shall set $\,G_4=1\,$).  Using these
equations and equation (\ref{brextcur}) with $\,T_{ij}=0\,$ in the
constraint equations (\ref{constr1}) and (\ref{constr2}) we find
that the momentum constraint equation is satisfied identically,
while the Hamiltonian constraint equation acquires the simple form
\begin{equation}
R=0\,\,, \label{ham1}
\end{equation}
which is in accord with equation (\ref{breq2}). Writing down this
equation explicitly in the metric (\ref{ansatz}) we obtain the
equation
\begin{equation}
\left(\frac{\partial ^2}{\partial r ^2} + \frac{4\,r}{\Sigma}\,
\frac{\partial}{\partial r} + \frac{2}{\Sigma} \right) H = 0\,\,.
\label{ham2}
\end{equation}
It can  easily be verified that this equation admits the solution of
the form
\begin{equation}
H  = \frac{2 M r}{\Sigma} - \frac{\beta}{\Sigma}\,\,\,, \label{sol1}
\end{equation}
where the parameters $\,M\,$ and $\,\beta\,$ are arbitrary constants
of integration. The physical meaning of these parameters becomes
more transparent when passing to the usual Boyer-Lindquist
coordinates in which the metric has more suitable asymptotic form.
Applying the Boyer-Lindquist transformation
\begin{eqnarray}
du &=& dt - \frac{r ^2 + a ^2}{\Delta}\,dr\,,
\nonumber \\[3mm]
d \varphi &=& d \phi - \frac{a}{\Delta}\,dr\,, \label{trans1}
\end{eqnarray}
where
\begin{equation}
\Delta = r^2 + a^2 - 2Mr + \beta\,, \label{delta}
\end{equation}
we find that the induced metric (\ref{ansatz}) with scalar function
(\ref{sol1}) can be brought to the form
\begin{eqnarray}
ds^2 &=& -\left(1 - \frac{2Mr - \beta}{\Sigma}\right)dt^2 -
 \frac{2 \,a\left(2Mr -\beta\right)}{\Sigma} \,\sin^2\theta \, dt\,d\phi
\nonumber\\
&& +\frac{\Sigma}{\Delta}\,dr^2\ + \Sigma \,d\theta^2 +
\left(r^2+a^2+ \frac{2Mr-\beta}{\Sigma}\,\,a^2 \sin ^2\theta
\right)\sin^2\theta \, d\phi^2\, \,.\label{kn}
\end{eqnarray}
We see that that this metric looks exactly like the familiar
Kerr-Newman solution in general relativity describing a stationary
and axisymmetric black hole with an electric charge. Since the
metric  is asymptotically flat, by passing to the far-field regime
one can interpret the parameter $\,M\,$  as the total mass of the
black hole and the parameter $\,a\,$ as the ratio of its angular
momentum to the mass. However, in our case there is no electric
charge on the brane, nevertheless  we have the parameter $\,\beta\,$
in the metric, which may take on both \textit{positive} and
\textit{negative} values.

From the asymptotic form of the metric (\ref{kn}) it follows that
the parameter $\,\beta\,$ in certain sense plays the role of a
Coulomb-type charge. On this ground, and following the case of
static black hole on the brane \cite{dmpr} we can think of it as
carrying the imprints of nonlocal Coulomb-type effects from the bulk
space, i.e. as a {\it tidal} charge parameter. In order to
strengthen this statement it would be instructive to calculate
explicitly the components of the tensor $\, E_{ij}\,$ on the brane.
Substituting the metric (\ref{kn}) into the field equation
(\ref{breq2}) we find the expressions
\begin{eqnarray}
{E _{t}}^t &=&- {E _{\phi}}^\phi=
 - \frac{\beta}{\Sigma^3} \left(\Sigma- 2\,(r^2+a^2)\right)\,,\nonumber\\
{E _{r}}^r &=&- {E _{\theta}}^\theta=
 \frac{\beta}{\Sigma^2}\,,\nonumber\\
{E _{\phi}}^t &=&- (r^2+a^2)\sin ^2 \theta \,{E _{t}}^\phi= -\frac{2
\beta a}{\Sigma ^3}\,(r^2+a^2)\,\sin ^2 \theta \,\,,
\label{eweylcomp}
\end{eqnarray}
which certainly obey the conservation condition $\,D_i E^i_j=0 \,$
on the brane. These quantities are close reminiscents of the
components of the energy-momentum tensor for a charged rotating
black hole in general relativity
\begin{eqnarray}
 {T _{t}}^t &=&- {T_{\phi}}^\phi=
 \frac{Q^2}{8 \pi \Sigma^3} \left(\Sigma- 2\,(r^2+a^2)\right)\,\,,\nonumber\\
{T _{r}}^r &=&- {T _{\theta}}^\theta= - \frac{Q^2}{8 \pi \Sigma^2}\,\,,\nonumber\\
{T _{\phi}}^t &=&- (r^2+a^2)\sin ^2 \theta \,{T _{t}}^\phi= \frac{
Q^2 a}{4 \pi \Sigma ^3}\,(r^2+a^2)\,\sin ^2 \theta\,\,,
\label{emtcomp}
\end{eqnarray}
provided that the square of the electric charge in the latter case
is "superceded" by the tidal charge parameter $\,\beta\,$.

In complete analogy to the Kerr-Newman solution in general
relativity, the metric (\ref{kn}) possesses two major features; the
event horizon structure and the existence of a  \textit{static
limit} surface. The event horizon is a null surface determined by
the equation $\,\Delta=0\,$. The largest root of this equation
\begin{equation}
r_{+}= M + \sqrt{M^2 - a^2 - \beta}\, \label{horizon1}
\end{equation} describes the position of the outermost event
horizon. We see that the horizon structure depends on the sign of
the tidal charge. The event horizon does exist provided that
\begin{equation}
 M^2 \geq a^2 +\beta\,,
\label{extreme}
\end{equation}
where the equality corresponds to the \textit{extreme} horizon. It
is clear that the positive tidal charge acts to weaken the
gravitational field and we have the same horizon structure as the
usual Kerr-Newman solution.

However, new interesting features arise when the tidal charge is
taken to be negative. Indeed, as it has been argued in a number of
papers (see Refs. \cite{chrss}\,, \cite{roy1}\,, \cite{sms2}) the
case of negative tidal charge is physically more natural one, since
it contributes to confining effect of the negative bulk cosmological
constant on the gravitational field in the RS scenario. In our case
the negative tidal charge leads to the possibility of a greater
horizon radius compared to the corresponding case of the Kerr-Newman
black hole. In particular, for $\, \beta < 0\,$ from equation
(\ref{horizon1}) one sees that as $\,a\rightarrow M\,$ the horizon
radius
\begin{equation}
r_{+} \rightarrow \left(M + \sqrt{ -\beta}\right)\, > M \,,
\label{greaterh}
\end{equation}
that is not allowed in the framework of general relativity.

On the other hand, as it follows from equations (\ref{horizon1}) and
(\ref{extreme}), for negative tidal charge the extreme horizon
$\,r_{+} = M \,$ corresponds to a black hole with rotation parameter
$\,a\,$ greater than its mass $\,M\,$\,. Thus, the bulk effects on
the brane \textit{may provide a mechanism for spinning up a
stationary and axisymmetric black hole on the brane so that its
rotation parameter exceeds its mass}. Meanwhile, such a mechanism is
impossible in general relativity.

The static limit surface is determined by the equation
$\,g_{tt}=0\,$, the largest root of which gives the radius of the
\textit{ergosphere} around  the black hole
\begin{equation}
r_{0} = M + \sqrt{M^2 - a^2 \cos^2\theta - \beta}\,\,. \label{ergo0}
\end{equation}
Clearly, this surface lies outside the event horizon coinciding with
it only at angles $\,\theta=0\,$ and $\,\theta=\pi\,$. The negative
tidal charge tends to extend the radius of the ergosphere around the
braneworld black hole, while the positive $\,\beta\,$, just as the
usual electric charge in the Kerr-Newman solution, plays the
opposite role. For the extreme case, taking equation (\ref{extreme})
into account in (\ref{ergo0}) we find the radius of the ergosphere
within
\begin{equation}
M<r<M+\sin\theta\,\sqrt{M^2-\beta}\,\,. \label{ergob}
\end{equation}
This once again confirms the statement made above about the
influence of negative and positive tidal charges on the ergosphere
region. We conclude that \textit{rotating braneworld black holes
with negative tidal charge are more energetic objects in the sense
of the extraction of the rotational energy from their ergosphere.}

\section{Rotating black hole with tidal and electric charges}

Let us now assume that the brane is not empty and there is a Maxwell
field trapped on the brane. Clearly, in this case a rotating black
hole on the brane may acquire an electric charge with respect to the
Maxwell field. In other words, rotating black holes on the brane
would in general carry both tidal and electric charges. Here we
shall study the exact metrics describing such black holes. Since the
trace of the energy-momentum tensor for a Maxwell field on the brane
vanishes and the bulk space is empty, the gravitational field
equations (\ref{breq1}) written in the RS model take the form
\begin{equation}
R_{ij} = 8 \pi T_{ij} + \kappa_5^4 \,\left(S_{ij} -
\frac{1}{2}\,h_{ij} S \right) - E_{ij}\,\,, \label{breq3}
\end{equation}
where the energy-momentum tensor of the electromagnetic field is
given by
\begin{equation}
T_{ij}= \frac{1}{4\,\pi} \left( F _{im} {F_j}^{\;m} - \frac{1}{4}\,
h_{ij}\, F _{lm} F ^{lm} \right)\,, \label{maxemt}
\end{equation}
the "squared" energy-momentum tensor
\begin{equation}
S_{ij} = -\frac{1}{4} \left( T_i^m\, T_{jm}- \frac{1}{2}\, h_{ij}\,
T_{lm} T ^{lm} \right)\,, \label{quademt3}
\end{equation}
and it is trace $\,S= S_{ij}\,h^{ij}\,$.

Following the line of the previous section we assume that the
induced metric on the non-empty brane still admits the Kerr-Schild
form given in (\ref{ansatz}), while the electromagnetic field on it
is described by a solution of source-free Maxwell equations . One
can easily verify that the solution of the Maxwell equations written
in terms of a potential one-form has the form
\begin{equation}
A = - \frac{Q r}{\Sigma} \left(du - a \sin ^2 \theta\, d\varphi
\right)\,, \label{potential}
\end{equation}
where the parameter $\,Q\,$ is thought of as the electric charge of
the black hole. With this potential one-form one can calculate the
non-vanishing components of the electromagnetic field tensor
$\,F_{ij}\,$ involved in (\ref{maxemt}) through the relation $\,
F=dA \,$. Having done this we find the expressions
\begin{eqnarray}
F _{ru} & =  & \frac{Q }{\Sigma^2} \left( r ^2 - a ^2 \cos ^2 \theta
\right)\,,~~~~~~~~~~~~~~~ F _{u\theta} = \frac{Q r a
^2\,\sin2\theta\
}{\Sigma^2}\,,~\nonumber\\[2mm]
F _{\varphi r} &=& \frac{Q a}{\Sigma^2}\,\left( r ^2 - a ^2 \cos ^2
\theta \right)\sin^2\theta\,\,,~~~~~~ F _{\theta \varphi} = \frac{Q
a r }{\Sigma ^2} \left( r ^2 + a ^2 \right) \sin 2\theta\,\,,
\label{covar}
\end{eqnarray}
while the contravariant components have the form
\begin{eqnarray}
F ^{ur} & =  & \frac{Q (r^2+a^2)}{\Sigma^3} \left( r ^2 - a ^2 \cos
^2 \theta \right)\,,~~~~~~ F ^{u\theta} = -\frac{Q r a
^2\,\sin2\theta\
}{\Sigma^3}\,,~\nonumber\\[2mm]
F ^{\varphi r} &=& \frac{Q a}{\Sigma^3}\,\left( r ^2 - a ^2 \cos ^2
\theta \right)\,\,,~~~~~~~~~~~~~~ F ^{\theta \varphi} = \frac{2\, Q
a r }{\Sigma ^3}\,\cot\theta\,\,.\label{contra}
\end{eqnarray}
Substituting these expressions into equation (\ref{maxemt}) we
obtain that the non-vanishing components of the energy-momentum
tensor precisely coincide with those of in (\ref{emtcomp}) taken at
$\,t\rightarrow u\,$ and $\,\phi \rightarrow \varphi $. Next, taking
this into account in equation (\ref{quademt3}) we calculate the
non-vanishing components of the "squared" energy-momentum tensor. We
find that
\begin{eqnarray}
{S_{u}}^u &=&
{S_{r}}^r={S_{\theta}}^\theta={S_{\varphi}}^\varphi=\frac{1}{4}\,S
\,\,, \label{quadcomp}
\end{eqnarray}
where the scalar $\,S \,$  is given by
\begin{equation}
S= \frac{Q^4}{ 64 \pi^2 \,\Sigma^4}\,\,. \label{trace}
\end{equation}

We turn now to the Hamiltonian constraint equation (\ref{constr1}),
which with the above equations in mind and using equations
(\ref{brextcur}) and (\ref{tuning}) can be brought to the form
\begin{equation}
R=-\kappa_5^4\,S\,. \label{ham3}
\end{equation}
The explicit form of this equation in the metric (\ref{ansatz})
gives the equation governing the scalar function $\,H\,$. We have
\begin{equation}
\left(\frac{\partial ^2}{\partial r ^2} + \frac{4\,r}{\Sigma}\,
\frac{\partial}{\partial r} + \frac{2}{\Sigma} \right) H=
-\frac{\ell^2 Q ^4}{\Sigma ^4}\,\,. \label{ham4}
\end{equation}
Solving this equation yields
\begin{equation}
H  =\frac{ 2 M r - Q^2- \beta}{\Sigma }  - \frac{l^2 Q
^4}{\Sigma}\,\, h\,\,, \label{gsolution}
\end{equation}
where the function $\,h\,$  is given by
\begin{equation}
h=\frac{1}{8 a ^4 \cos ^4 \theta} \left[ 2 + \frac{ r ^2 }{ \Sigma }
+ \frac{ 3 r }{a \cos \theta} \arctan\left(\frac{ r }{ a \cos \theta
} \right) \right] \,\,.\label{h}
\end{equation}
It should be noted that in obtaining this solution two arbitrary
functions of $\,\theta \,$ alone arising in the integration
procedure were specified from the correct far-field limit of the
solution. It is clear that substituting the metric (\ref{ansatz})
with (\ref{gsolution}) into the field equations (\ref{breq3}) on the
brane and  taking into account equations (\ref{covar})-(\ref{trace})
one can calculate explicitly the "electric" part $\,E_{ij}\,$ of the
bulk Weyl tensor through straightforward calculations, thereby
closing the system of field equations on the brane.

To explore the further properties of  metric (\ref{ansatz}) with
scalar function (\ref{gsolution}), as in the case of electrically
neutral rotating black hole considered in the previous section, it
is useful to put this metric in the Boyer-Lindquist form as well.
However, it  does not admit the Boyer-Lindquist form in the usual
sense, where the only off-diagonal component of the metric with
indices $\,t\,$ and $\,\phi\,$ survives. Instead, we also have the
off-diagonal components of the metric with indices $\,(r \phi)\,$
and $\,(r t)\,$ that arise due to the combined influence of local
bulk effects on the brane and the \textit{dragging effect} of the
rotating black hole.

In a  way similar to (\ref{trans1}), one can always consider a
Boyer-Lindquist type transformation provided that the latter
transforming the metric, at the same time leaves  equation
(\ref{ham4}) unchanged. The transformation that fulfils this
requirement has the form
\begin{eqnarray}
du &=& dt - \frac{r ^2 + a ^2}{\Delta_{\ell}}\,dr\,,
\nonumber \\[3mm]
d \varphi &=& d \phi - \frac{a}{\Delta_{\ell}}\,dr\,, \label{trans2}
\end{eqnarray}
where
\begin{equation}
\Delta_{\ell}= r^2 + a^2 - 2Mr + \beta +Q^2 +\ell^2 Q^4\, h_0
\label{delta1}
\end{equation}
and $\,h_0\,$ is given by (\ref{h}) taken at a fixed angle
$\,\theta_0\, .$ Under this transformation the metric (\ref{ansatz})
with (\ref{gsolution}) takes the form
\begin{eqnarray}
ds^2 &=& -\left(1 - H \right)dt^2
+\frac{\Sigma}{\Delta_{\ell}}\,\left(1+\delta \right)dr^2\ + \Sigma
\,d\theta^2 +\left(r^2+a^2+ H a^2 \sin^2\theta\right)\sin^2\theta \,
d\phi^2
\nonumber\\[2mm]
&&-2\, \delta \left(dt-a \sin^2\theta \,d\phi\right)dr-2\,a H
\sin^2\theta \,dt\,d\phi\,\,,
 \label{bltmetric}
\end{eqnarray}
where
\begin{equation}
\delta= -\frac{\ell^2 Q^4}{\Delta_{\ell}}\,(h-h_0)
\,\,.\label{cross}
\end{equation}

At large distances the metric (\ref{bltmetric}) approaches the
Lense-Thirring form, thereby confirming that it describes a slowly
rotating body on the brane. It can be easily shown that for
$\,a=0\,$ this metric goes over into the metric of a charged static
braneworld black hole obtained in \cite{chrss}, while for $\,Q=0\,$
we have the metric (\ref{kn}) for a rotating black hole with tidal
charge. Indeed, expanding the solution (\ref{gsolution}) in powers
of the rotation parameter $\,a\,$ and keeping the correction terms
up to the second order, we obtain the expressions
\begin{equation}
H =  \frac{ 2M }{ r} -\frac{\beta + Q^2}{r^2} - \frac{ l^2 Q^4 }{
20\, r ^6}\;- \;\frac{ a^2 \cos^2\theta}{r^2}\left( \frac{2M}{r} -
\frac{ \beta + Q^2 }{r^2} -\frac{ 17 }{ 140 } \frac{ l^2 Q^4 }{ r^6
}\, \right)\,\,, \label{expan1}
\end{equation}

\begin{equation}
h-h_0=-\frac{1}{14}\,\frac{a^2}{r^6}\,\left(\cos^2\theta-\cos^2\theta_0\right)\,.
\label{expan2}
\end{equation}
Taking these expansions into account in the metric
(\ref{bltmetric}), for $\,a\rightarrow 0\,$ we arrive at the
statement made above. Furthermore, we see that the off-diagonal
components of the metric involving an index $\,r\,$ are proportional
only to the square of the rotation parameter. Thus, for small enough
values of the rotation parameter (i. e. in linear in $\,a\,$
approximation) the metric takes the usual Boyer-Lindquist form and
describes a charged and slowly rotating black hole on the brane.
However, in the case of arbitrary rotation the metric
(\ref{bltmetric}) contains extra off-diagonal components determined
by the composite influence of the local bulk effects on the brane
and the rotational dynamics of the black hole.

We also note that our solutions, both (\ref{kn}) and
(\ref{bltmetric}), do not match the far distance behavior of the
"black hole metric" on the Randall-Sundrum brane computed in
\cite{gt}-\cite{gkr} within a linear perturbation analysis. The
latter contains  $\,r^{-3}\,$ corrections to the metric which are
absent in our solutions (see Eq.(\ref{expan1})). However, as it has
been argued in \cite{chrss}, for black holes of mass $\,M\gg
\ell\,,$ the post-Newtonian corrections to the metric dominate over
the RS-type corrections, therefore the comparison of our results
with that of the linearized analysis is not appropriate beyond the
leading order correction terms. On the other hand, our solutions
may, to a good enough accuracy, describe the strong gravity regime
at short distances on the brane, at which the dominating behavior of
the gravitational potential becomes like $\,r^{-2}\,$ \cite{dmpr}.

Another feature of the expression (\ref{expan1}) is that it involves
the tidal charge parameter $\,\beta\,$ and the square of the
electric charge $\,Q^2\,$ at the same footing as two Coulomb-type
terms at large distances. This, in a certain sense, again confirms
the interpretation of $\,\beta\,$ as a tidal charge. From
(\ref{expan1}) it also follows that the contributions of the tidal
and electric charges to the Coulomb field of the black hole may
precisely cancel each other provided that $\,\beta=- Q^2\,\,$ thus,
leaving us only with higher order correction terms.

It is also worthy of noting that in addition to transformation
(\ref{trans2}) there exists a second transformation
\begin{eqnarray}
du &=& dt - \frac{r ^2 + a
^2}{\Delta_{\ell}}\,\left(\frac{1+\delta}{1-\delta}\right)dr\,,
\nonumber \\[3mm]
d \varphi &=& d \phi -
\frac{a}{\Delta_{\ell}}\,\left(\frac{1+\delta}{1-\delta}\right)
dr\,, \label{trans3}
\end{eqnarray}
which brings the metric (\ref{ansatz}) with (\ref{gsolution}) into
the form
\begin{eqnarray}
ds^2 &=& -\left(1 - H \right)dt^2
+\frac{\Sigma}{\Delta_{\ell}}\,\left(1+\delta \right)dr^2\ + \Sigma
\,d\theta^2 +\left(r^2+a^2+ H a^2 \sin^2\theta\right)\sin^2\theta \,
d\phi^2
\nonumber\\[2mm]
&&+2\, \delta \left(dt-a \sin^2\theta \,d\phi\right)dr-2\,a H
\sin^2\theta \,dt\,d\phi\,\,.
 \label{bltmetric1}
\end{eqnarray}
We see that this metric differs from the metric (\ref{bltmetric})
only by the opposite sign of the off-diagonal components  with
indices $\,(r \phi)\,$ and $\,(r t)\,$  and these two metrics go
over into each other under the transformation
$\,t\rightarrow-t\,\,,\,\,\,\,\phi\rightarrow-\phi\,\,.$  We also
emphasize that there are no other transformations of types
(\ref{trans2}) and (\ref{trans3}) which put the metric
(\ref{ansatz}) in the Boyer-Lindquist type form along with
preserving equation (\ref{ham4}).

Turning back to the metric (\ref{bltmetric}) we see that its
components involving an index $\,r\,$ become singular at
$\,\Delta_{\ell}=0 \,$, where the event horizon of the black hole
occurs. Explicitly we have the following equation governing the
radius of the event horizon
\begin{equation}
r ^2 + a ^2 - 2Mr + \beta + Q^2 + \frac{ l^2 Q^4 }{8 a ^4 \cos ^4
\theta_0 } \left[ 2 + \frac{ r ^2 }{ \Sigma_0 } + \frac{ 3 r }{ a
\cos \theta_0 } \arctan \left( \frac{ r }{ a \cos \theta_0 } \right)
\right] = 0 \,\,. \label{horizon}
\end{equation}

First of all, from this equation it follows that if the rotation of
the black hole on the brane occurs slowly enough $\,(a\ll M)\,$, the
horizon structure is similar to that of a slowly rotating charged
black hole in general relativity. However, for arbitrary rotation
the situation is drastically changed and in this case equation
(\ref{horizon}) indeed defines a null orbit for each fixed
$\,\theta\,.$ In order to show this, in a similar way to the
description of the horizon of stationary black holes in general
relativity \cite{carter1}, we suppose that the isometry of the event
horizon is described by a Killing vector field, which must be
tangent to the null orbits of the horizon. The Killing vector can be
constructed as a linear combination of the two commuting, the
timelike and the rotational Killing vectors
\begin{eqnarray}
\,\xi_{(t)}&= &\frac{\partial}{\partial t}\,\,,~~~~~~~~~
\,\xi_{(\phi)}=\frac{\partial}{\partial \phi}\,\,, \label{killings}
\end{eqnarray}
of the metric (\ref{bltmetric}) as follows
\begin{equation}
\chi=\xi_{(t)} + \Omega\,\xi_{(\phi)}\,\,, \label{hkilling}
\end{equation}
where the quantity
\begin{equation}
\Omega=-\frac{g_{t\phi}}{g_{\phi \phi}} \label{angular}
\end{equation}
is the angular velocity of "locally non-rotating observers" orbiting
around the black hole at fixed values of $\,r\,$ and $\,\theta\,$.
One can easily verify that the vector defined in (\ref{hkilling})
becomes null at  $\,\Delta_{\ell}=0\,\,$ i.e. it is tangent to the
null orbit of the horizon at fixed $\,\theta\,$. At this orbit the
angular velocity of the rotation takes the form
\begin{equation}
\Omega_H=\frac{a}{r_{+}^2+a^2} \,\,. \label{hangular}
\end{equation}
This quantity is a close reminiscent of its counterpart for a
stationary black hole in general relativity, however, as it is seen
from equation (\ref{horizon}), here the event horizon is determined
for fixed  values of $\,\theta $ and, therefore the black hole
undergoes differential rotations with non-uniform angular velocities
at different $\,\theta $-layers. Roughly speaking, one can think of
the horizon structure of the black hole as composed of null circles
extending from the equatorial plane to the poles. Each of these
circles rotates with its own angular velocity. Thus, \textit{a
rotating braneworld black hole carrying both types of charges in
general has a differentiated horizon structure for each values of
the $\,\theta$-angle.} From (\ref{bltmetric}) and (\ref{bltmetric1})
we see that at the horizon when $\,\theta\rightarrow \theta_0\,$,
indeed the off-diagonal components with indices $\,(r \phi)\,$ and
$\,(r t)\,$ tends to vanish, in accordance with the physical meaning
of any rotation.

We have applied numerical calculations to analyze the largest roots
of equation (\ref{horizon}). The numerical results are plotted in
Figs. 1(a,b) and Fig. 2. For the sake of certainty in all numerical
calculations we have set the curvature radius of AdS $\,l=0.01
M^2\,. $ In Fig. 1(a) the plots illustrate the dependence of the
horizon radius on the electric charge of the black hole with
vanishing rotation parameter $(\,a=0\,)$ and different values of the
tidal charge $\,\beta=0\,,\,\, \pm 0.5 M^2\,,\,-5 M^2\,,\,\,-20
M^2\,\,$. As expected, the size of the horizon is sensitive to the
sign of tidal charge. We see that, in contrast to its  positive
values, the negative tidal charge has greatest effect of increasing
the horizon radius, while the electric charge of the black hole
opposes it. In all cases the horizon radius  decreases with
increasing  electric charge and the critical values of the electric
charge at which the horizon does exist are different for different
values of $\,\beta\,.$ Fig. 1(b) involves the similar plots with the
value of the rotation parameter $\,a=0.5 M^2\,$ in the equatorial
plane $(\,\theta=\pi/2\,).$ One sees that the effect of rotation of
the black hole, just as that of electric charge, is to decrease the
horizon radius.

Fig. 2  presents some results of  the numerical analysis of the
horizon radius for a particular value of the tidal charge $\,
\beta=-Q^2\,.$ The plots reveal the different size of the horizon in
different $\,\theta\,$-planes with increasing absolute value of the
tidal charge. The effect becomes more significant with the growth of
rotation parameter. We see that for given charge the horizon radius
at poles $\,\theta=0\,\,,\,\pi\,$ is always greater than at
$\,\theta=\pi/2\,$ . Furthermore, the critical absolute value of
$\,\beta\,$ at which the horizon exist is also greater for
$\,\theta=0\,\,,\,\pi\,.$

To conclude this section we note that the boundary of the ergosphere
in metric (\ref{bltmetric})can  be described in a usual manner, that
is by the condition $\, g_{tt}=0\,$. This can be also written in the
form
\begin{equation}
r ^2  - 2Mr + \beta + Q^2 + a ^2 \cos^2 \theta + \frac{ l^2 Q ^4 }{8
a ^4 \cos ^4 \theta } \left[ 2 + \frac{ r ^2 }{ \Sigma } + \frac{ 3
r }{ a \cos \theta } \arctan \left( \frac{ r }{ a \cos \theta }
\right) \right] =0 \,\,.\label{ergosphere}
\end{equation}
The comparison of this equation with (\ref{horizon}) shows that the
ergosphere region always lies beyond the event horizon coinciding
with it only at poles $\,\theta=0\,,\,\,\pi\,$.

\section{Geodesic motion}

It is generally believed that in situations of astrophysical
interest the electric charge of a Kerr black hole will quickly
become negligible due to the neutralizing effect of an ionized
medium surrounding the black hole. The same argument may remain true
for a rotating black hole in the braneworld scenario, where charged
particles can live only on the brane and their selective accretion
by the black hole will significantly diminish its electric charge.
However, in the latter case even without electric charge being
present, the black hole may still have a tidal charge emerging as a
pure geometrical effect from the bulk space \cite{dmpr}. Therefore,
the above argument does not apply to constraint the value of the
tidal charge and it, in principle, may have its greatest effect on
physical processes in the strong-gravity region around the black
hole. To get some insights into this we shall now study the geodesic
motion of test particles in the metric (\ref{kn}) of a rotating
black hole with tidal charge .

We start with the Hamilton-Jacobi equation
\begin{equation}
h^{ij}\,\frac{\partial S}{\partial x^{i}} \,\frac{\partial
S}{\partial x^{j}} + m^2 =0, \label{hj}
\end{equation}
where $m$ is the mass of a test particle. Following Carter's result
\cite{carter} of the complete separability of this equation in the
metric of a Kerr-Newman black hole in general relativity we can
write the action in the form
\begin{equation}
S=-E t + L \phi + S_{r}(r) + S_{\theta}(\theta) \,, \label{action}
\end{equation}
where the conserved quantities $E$ and $L$ are the energy and the
angular momentum of the test particle at infinity, respectively.
Substituting it into equation (\ref{hj}) we obtain the two equations
in the separable form
\begin{eqnarray}
\Delta \left(\frac{d S_r}{d r}\right)^2
-\frac{1}{\Delta}\,\left[(r^2+a^2)E -a \,L\right]^2+ m^2 r^2 &= &
-\textsc{K} \,\,,
\label{sepeq1}\\[2mm]
\left(\frac{d S_{\theta}}{d \theta}\right)^2 + \left(a E\sin \theta-
\frac{L}{\sin\theta} \right)^2+ m^2 a^2 \cos^2\theta &= &
\textsc{K}\,\, ,\label{sepeq2}
\end{eqnarray}
where $\,\textsc{K}\,$ is a constant of separation. We shall
restrict ourselves to geodesic motion in the equatorial plane
$\,\theta=\pi/2\,$. Then using equations (\ref{sepeq1}) and
(\ref{sepeq2}), along with the action(\ref{action}), in the
expression
\begin{equation}
\frac{dx^i}{d\lambda}= h^{ij}\,\frac{\partial S}{\partial x^{j}}
\,\,,
\end{equation}
where  $\,\lambda= s/m\,$ is an affine parameter along the
trajectory of the particle, we arrive at the following equations of
motion
\begin{eqnarray}
\Delta\,\frac{d t}{d \lambda}&=&\left(r^2+a^2+\frac{2 M
r-\beta}{r^2}\,\,a^2\right) E- \frac{2 M r-\beta}{r^2}\,\,a
L\,\,,\nonumber \\[3mm]
\Delta\,\frac{d \phi}{d \lambda}&=& \left(1-\frac{2 M
r-\beta}{r^2}\right)L+\frac{2 M r-\beta}{r^2}\,\,a E \,\,,
\label{time}
\end{eqnarray}
and
\begin{equation}
r^4\,\left(\frac{dr}{d\lambda}\right)^2 = V(E,L,r,a,\beta)\,\,,
\label{radial}
\end{equation}
where the effective potential of the radial motion is given by
\begin{eqnarray}
V=\left[(r^2+a^2)\,r^2+(2 M r -\beta)\,a^2\right] E^2 -(r^2-2 M r
+\beta)\,L^2
\nonumber \\
 - 2\,a (2 M r-\beta)\,E  L - m^2 r^2 \Delta\,\,.
 \label{effpot}
\end{eqnarray}
From the symmetry of the problem it follows that a simple example of
the geodesic motion, the circular motion, occurs  in the equatorial
plane of the black hole. The energy and the angular momentum of the
circular motion at some radius are determined by the simultaneous
solution of the equations
\begin{eqnarray}
V&=& 0\,\,,~~~~~~~~~~\frac{\partial V}{\partial r} =0 \,\,.
\label{eq1}
\end{eqnarray}
Solving these equation yields
\begin{equation}
\frac{E}{m}= \frac{r^2-2 M r+\beta \pm a\, \sqrt{Mr-\beta}}
{r\,\left[r^2-3 M r+2 \beta \pm 2\, a \,\sqrt{M
r-\beta}\,\right]^{1/2}}\,\,\,, \label{energy}
\end{equation}
\vspace{3mm}
\begin{equation}
\frac{L}{m}= \pm\,\frac{\sqrt{Mr-\beta}\left(r^2+a^2 \mp 2\,a\,
\sqrt{Mr-\beta}\right) \mp a \beta} {r\,\left[r^2-3 M r+2 \beta \pm
2\, a \,\sqrt{M r-\beta}\,\right]^{1/2}}\,\,\,.\label{momentum}
\end{equation}
Here and in the following the upper sign corresponds to the
so-called direct orbits in which the particles are corotating $\, (L
>0 \,)$ with respect to the rotation of the black hole, while the
lower sign applies to the retrograde, counterrotating $\, (L < 0)\,$
motion of the particles.

From equation (\ref{energy}) it follows that the region of existence
of the circular orbits extends from infinity up to the radius of the
limiting photon orbit, which is determined  by the condition
\begin{equation}
r^2-3 M r +2 \beta \pm 2\, a \sqrt{M r-\beta}=0 \,\,.\label{exeq}
\end{equation}
It is evident that in the region of existence not all circular
orbits are bound. The radius of the marginally bound orbits with
$\,E^2=m^2\,$ is given by the largest root of the polynomial
equation
\begin{equation}
M r \left(4 M r-r^2 -4 \beta +a^2 \right)+ \beta
\left(\beta-a^2\right) \pm\,2 a \left(\beta-2 M r\right)\sqrt{M
r-\beta}=0 \,\,.\label{boundeq}
\end{equation}
The condition for stability of the circular motion is given by the
inequality
\begin{equation}
\frac{\partial^2 V}{\partial r^2} \leq 0 \,\,,
\label{stability}
\end{equation}
where the case of equality corresponds to the marginally stable
circular orbits. Next, substituting  expressions (\ref{energy}) and
(\ref{momentum}) into (\ref{stability}) we obtain the equation
governing the boundary of the marginally stable orbits
\begin{equation}
M r \left(6 M r-r^2 - 9 \beta + 3 \,a^2 \right)+ 4 \beta
\left(\beta-a^2\right) \mp \,8  a \left(M
r-\beta\right)^{3/2}=0\,\,.
\label{stabeq}
\end{equation}

We note that the similar equations of the form of (\ref{exeq}),
(\ref{boundeq}) and (\ref{stabeq}) for a circular geodesic motion in
the equatorial plane of the Kerr-Newman black hole in general
relativity were obtained a long time ago in \cite{dk}. The numerical
analysis of the boundaries of the equatorial circular orbits in the
Kerr-Newman metric as functions of the rotation parameter and the
electric charge of the black hole was given in \cite{alievg}. In
particular, it has been shown that the radius of the limiting photon
orbit, as well as the radii of the innermost bound and the innermost
stable circular orbits moves towards the event horizon with the
growth of black hole's electric charge both for direct and
retrograde motions.

It is evident that in our case the positive tidal charge will play
the same role in its effect on the circular orbits as the electric
charge in the Kerr-Newman metric. However, as it has been emphasized
in Sec.III, the case of negative tidal charge is physically more
natural case on the brane and therefore, here we wish to know how
the negative tidal charge affects the circular geodesics. For this
purpose, we apply a numerical analysis to study the solutions of
equations (\ref{exeq}), (\ref{boundeq}) and (\ref{stabeq}). The
numerical analysis shows  that, regardless to the direct, or
retrograde motions of the particles, the limiting radii for
existence, bound and stability of the circular orbits enlarges from
the event horizon as the absolute value of $\,\beta \,$ increases.
In particular, for $\,\beta=-M^2 \,$ and $\,a=0\,$ we find that the
radii of the limiting photon and the last stable orbits are given as
\begin{eqnarray}
r_{ph}&\simeq &3.56 M\,,~~~~~~r_{ms}\simeq 7.3 M \,,
\label{example1}
\end{eqnarray}
while the radius of the event horizon $\,r_{+}=(1+\sqrt{2}) M \,.$
For a comparison we recall that for the same but positive value of
the tidal charge, which at the same time is the limiting value of
the charge in the Reissner-N\"{o}rdstrom metric one finds
\begin{eqnarray}
r_{ph}&= & 2 M\,,~~~~~~r_{ms}=4 M \,, ~~~~~~r_{+}=M\,.
\label{example2}
\end{eqnarray}
For these two cases it is also instructive to compare the binding
energies per unit mass of a particle at the last stable circular
orbits. Thus, for $\,\beta=-M^2 \,$ using (\ref{energy}) we find
that
\begin{eqnarray}
E_{binding}=1-\frac{E}{m}\simeq4.7\,\%\,\,,
\end{eqnarray}
in contrast to the binding energy $\,E_{binding}\simeq 8.1 \,\%\,\,$
of a particle in the case of $\,\beta=M^2 \,.$ We conclude that in
the $\,\beta < 0 \,$ case the binding energy of a particle moving in
the last stable circular orbits around a static black hole
\textit{decreases} in contrast to the $\,\beta > 0 \,$ case, where
it increases attaining its  maximum value  for the limiting case of
$\,\beta=M^2 \,.$

For a rotating black hole with the negative tidal charge
$\,\beta=-M^2 \,$ and  the rotation parameter $\,a=M \,,$ the
numerical calculations for the direct motion give
\begin{eqnarray}
r_{ph}&\simeq & 2.89 M\,,~~~~~~r_{ms}\simeq 3.89 M
\,,~~~~~E_{binding}\simeq 9\,\%\,,\label{example3}
\end{eqnarray}
while, for the retrograde motion we obtain
\begin{eqnarray}
r_{ph}&\simeq & 4.48 M\,,~~~~~~r_{ms}\simeq 10.15 M
\,,~~~~~E_{binding}\simeq 3\,\%\,.\label{example4}
\end{eqnarray}
We recall that for this case the pertaining radius of the horizon
radius $\,r_{+}=2 M \,.$ In the meantime, the counterparts of these
expressions \cite{bardeen} for a maximally rotating black hole in
general relativity $(\,a\rightarrow M \,,\,\,\,r_{+}\rightarrow M
\,)$ are given by $\, r_{ph}\simeq M\,,~~r_{ms}\simeq  M
\,,~~E_{binding}\simeq 42\,\%\,$ for the direct orbit, and
$\,r_{ph}\simeq  4 M\,,~~r_{ms}\simeq 9 M\,,~~E_{binding}\simeq
3.7\,\%\,$ for the retrograde  orbit. We note that compared to the
case of a maximally rotating black hole in general relativity for
which only zero-electric charge is acceptable, the presence of
negative tidal charge increases the radii of the marginally stable
orbits in both direct and retrograde motions, thereby decreasing the
corresponding binding energies. However, it should be emphasized
that the value of the rotation parameter $\,a=M \,$ is not a
limiting value for the braneworld black hole carrying negative tidal
charge. For instance, for $\,\beta=-M^2 \,$ it becomes $\,a=\sqrt{2}
M \,.$ In this case, the radius of the horizon is $\,r_{+}= M \,$,
and instead of the expressions (\ref{example3}) and (\ref{example4})
we have
\begin{eqnarray}
r_{ph}&\simeq &  M\,,~~~~~~r_{ms}\simeq  M \,,~~~~~E_{binding}\simeq
46.5\,\%\,\,, \label{example7}
\end{eqnarray}
for the last direct circular orbit and
\begin{eqnarray}
r_{ph}&\simeq & 4.82 M\,,~~~~~~r_{ms}\simeq 11.25
M\,,~~~~~E_{binding}\simeq 3\,\%\, , \label{example8}
\end{eqnarray}
for the last retrograde one.  In particular, from equation
(\ref{example7}) we see that for the class of direct orbits, the
negative tidal charge tends to \textit{increase} (through increasing
the limiting value of the rotation parameter compared to that of in
general relativity) the efficiency of an accretion disc around a
maximally rotating braneworld black hole. The details of the
numerical results are plotted in Figs. 3(a,b) and 4(a,b) . In figure
3(a) for  the $\,\beta<0\,$ case the curves display the growth of
the horizon size, the radius of the limiting photon orbit, as well
as the radii of the last bound and  the last stable circular orbits
with increasing absolute value of the tidal charge of a non-rotating
black hole. For the same case of $\,\beta<0\,$, the figure 3 (b)
plots the same curves for given non-zero value of black hole's
rotation parameter $\,a=M\,.$  We see that for all cases increasing
the absolute value of the tidal charge appears to \textit{increase}
the radii of the limiting circular orbits.

The curves of figure 4 (a,b) illustrate the dependence of the
horizon size, the radius of the limiting photon orbit and the radii
of the marginally bound and the marginally  stable orbits on the
rotation parameter of the black hole for given $\,\beta=\pm \,0.5
M^2\,.$  The positive tidal charge acts  to \textit{decrease} the
radii of circular orbits in both direct and retrograde motions, just
as an electric charge in general relativity (Fig.4(a)). On the other
hand, the negative tidal charge has its opposite effect of
\textit{increasing} the radii, regardless to a particular class of
circular orbits (Fig. 4(b)). In both cases the effect of the
rotation is to decrease the radii for direct orbits and to increase
them for retrograde ones.

\section{Conclusion}

The exact bulk metrics that consistently describe the formation of
black holes as endpoints of gravitational  collapse of matter on a
3-brane in the Randal-Sundrum scenario have not been constructed
yet. One of the attempts made in this direction is based on a
specialization of the induced metrics on the brane and  then solving
the effective gravitational field equations for these metrics. The
subsequent integration of the field equations into the bulk using
the induced metrics as "initial data"  could provide the desired
solutions for braneworld black holes. Following the similar line, in
this paper we have specified a particular metric ansatz for rotating
black holes localized on a 3-brane in the Randall-Sundrum braneworld
. Namely, we have assumed that the metrics describing rotating black
holes on the brane would admit the familiar Kerr-Schild form.

First, we have considered the case of an empty brane and solved the
constraint equations on the brane, which together with the effective
gravitational field equations on the brane form a closed set of
equations for our Kerr-Schild ansatz. We have shown  that the
resulting metric is not in a usual Kerr form, instead it involves an
extra non-trivial parameter which looks like to play exactly the
same role as a Maxwell electric charge in the Kerr-Newman solution
of general relativity. However, the extra parameter in our case has
a pure geometrical origin and can be naturally interpreted as a
\textit{tidal} charge transmitting to the brane the gravitational
imprints of the bulk space. It is interesting to note that, in
contrast to the "squared" electric charge in the Kerr-Newman metric,
the tidal charge in the rotating black hole metric on the brane may
take on both positive and negative values. Moreover, as it has been
argued in a number of papers (see Refs. \cite{chrss}, \cite{roy1},
\cite{sms2}) that the case of negative tidal charge is physically
more natural one, since it acts in the same direction as the
negative cosmological constant to keep the gravitational field (the
horizon of a black hole) around the brane.

We have also studied the effects of negative tidal charge on the
horizon structure and the ergosphere of the rotating black hole on
the empty brane. In particular, we have found that the negative
tidal charge may provide a mechanism for spinning up the black hole
to the value of its rotation parameter which is greater than its
mass. This is not allowed in the framework of general relativity.
The negative tidal charge extends the boundary of the ergosphere,
thereby making the rotating braneworld black hole a more energetic
object with respect to the extraction of its rotational energy.

Next, we have turned to the case of non-empty brane, assuming the
presence of a Maxwell field on it. In this case a rotating black
hole on the brane, in addition to its tidal charge, would also carry
an electric charge of the Maxwell field. Using again the Kerr-Schild
ansatz for the induced metric on the brane  we have presented a new
solution describing such a black hole . It turned out that for a
rapid enough rotation the composite effects of the black hole
rotational dynamics and the "squared" energy-momentum tensor on the
brane distort the geometry of the black hole in such a way that it
has a differentiated horizon structure at each fixed
$\,\theta$-plane. In other words, the horizon depends on the values
of $\,\theta$-angle, and therefore the horizon structure of the
black hole can be thought of as composed of non-uniformly rotating
null circles. We have also numerically solved the corresponding
equation governing the radii of the circles. In particular, we have
shown that for given special values of the charges the radius of the
horizon at poles $(\,\theta\,=0\,,\,\pi )$ is greater than that of
in the equatorial plane $(\,\theta =\,\pi/2 )\,.$ The difference
becomes more significant with the growth of black hole's rotation
parameter.

Finally, we have investigated the circular geodesic motion of test
particles in the equatorial plane of a rotating braneworld black
hole carrying a tidal charge. We have found that the negative tidal
charge has its greatest effect in increasing the radius of the
horizon, as well as the radii of existence, boundedness and
stability for the innermost circular orbits, regardless to the
direct, or retrograde classes of the orbits.

We would like to note that it is not clear that the Kerr-Schild
ansatz for the induced metrics on the brane we have used in this
paper to describe rotating braneworld black hole is indeed fulfilled
by an exact bulk metric. Nevertheless, we believe that our approach
even without constructing exact bulk metrics gives significant
insight into the physics of rotating black holes localized on a
3-brane in the Randall-Sundrum scenario. In the context of possible
observational signatures of the rotating black holes with tidal
charge it would be interesting to study their gravitational lensing
effects. Certainly, it would be also very interesting to explore the
evolution into the bulk space of the black hole solutions given in
this paper using both analytical and numerical calculations.

\section{Acknowledgment}

We would like to thank Roy Maartens for useful discussions at an
early stage of this work.

\begin{figure}
\includegraphics[width=10cm,clip]{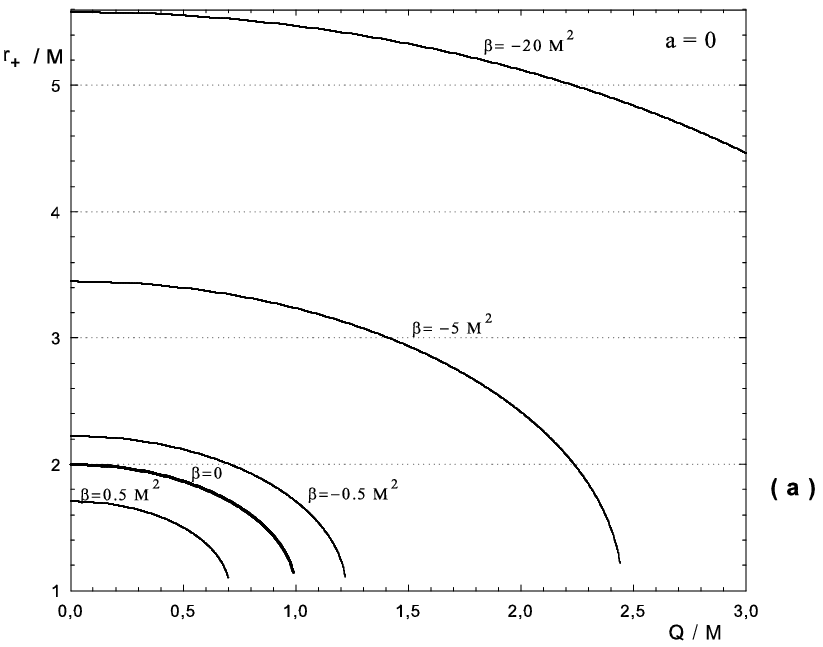}

~

~

\includegraphics[width=10cm,clip]{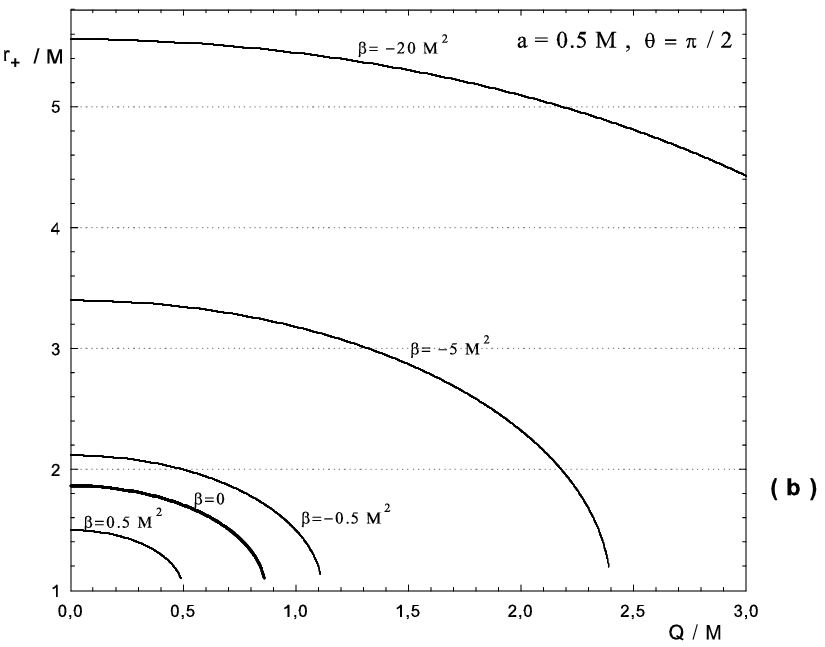}
\caption{ \label{horizon,fixed a} The dependence of the size of the
horizon of black holes localized on a 3-brane in the Randall-Sundrum
braneworld on the electric charge of the black holes for different
values of their tidal charge $\,\beta\, .$ Figure (a) corresponds to
a non-rotating black hole, while figure (b) is for a rotating black
hole with $\,a=0.5 M \,$and  $\,\theta=\pi/2\,$. }
\end{figure}
\newpage

\begin{figure}
\vspace{50mm}

\includegraphics[width=9.4cm,clip]{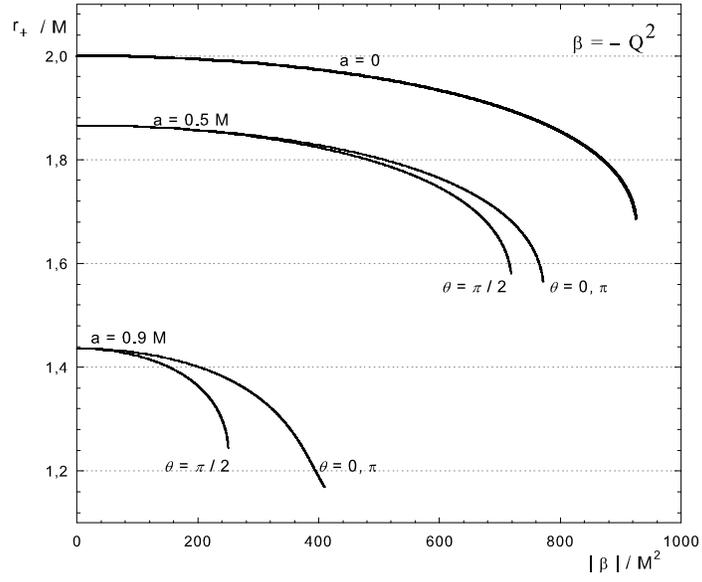}
\caption{ \label{horizon, beta=-Q^2} The dependence of the horizon
radius of a rotating braneworld black hole in its equatorial plane
$\,\theta=\pi/2\,$ and at poles $\,\theta=0\,,\,\pi\,$ on the
absolute value of its tidal charge $\,\beta=-Q^2\,$ for given values
of the rotation parameter $\,a=0\,,\,\,0.5 M\,,\,\, 0.9 M\, .$ }
\end{figure}
\newpage

\begin{figure}
\includegraphics[width=10cm,clip]{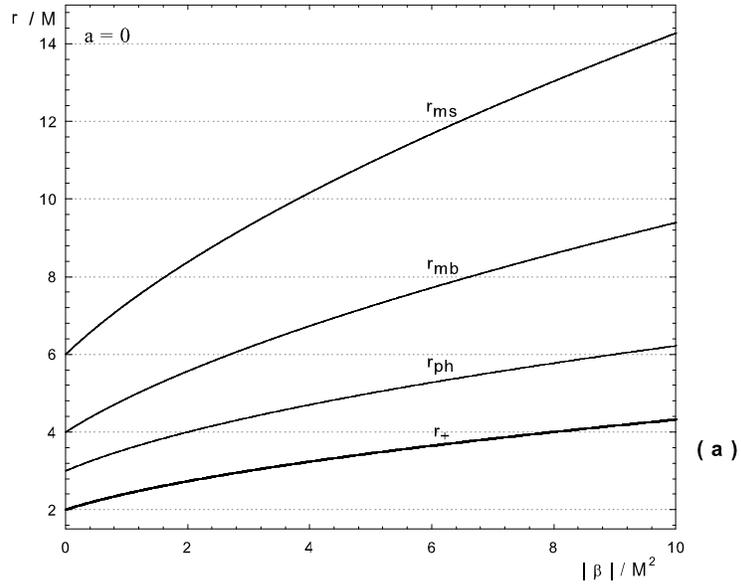}

~

~

\includegraphics[width=10cm,clip]{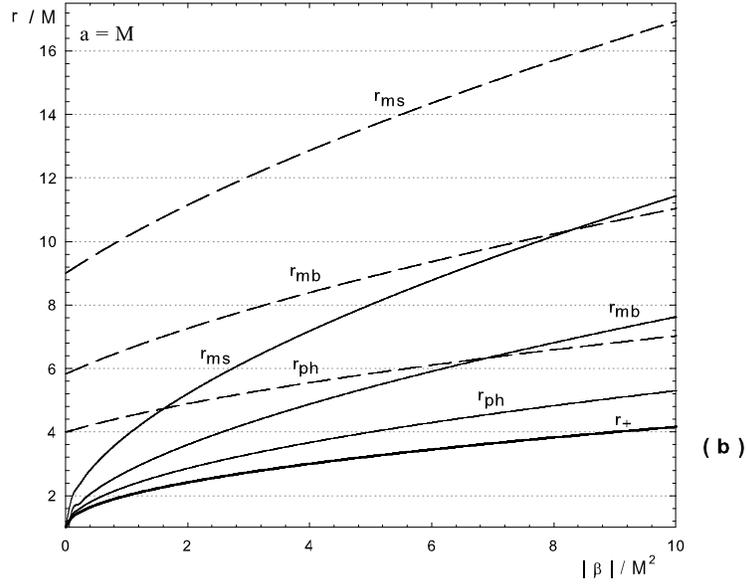}
\caption{ \label{geodesics,fixeda} Radii of the event horizon, the
limiting photon orbit, the marginally bound and the marginally
stable circular orbits in the equatorial plane $(\,\theta=\pi/2\,)$
of a rotating black hole carrying a tidal charge. The curves
illustrate the dependence of the radii on the absolute value of the
tidal charge $(\,\beta<0\,)$ for given values of the rotation
parameter $\,a=0\,$ (figure (a)) and $\,a=M\,$ (figure (b)). The
solid curves correspond to direct orbits, whereas the dashed curves
refer to retrograde orbits. The position of the horizon is indicated
by the bold curve.}
\end{figure}
\newpage

\begin{figure}
\includegraphics[width=10cm,clip]{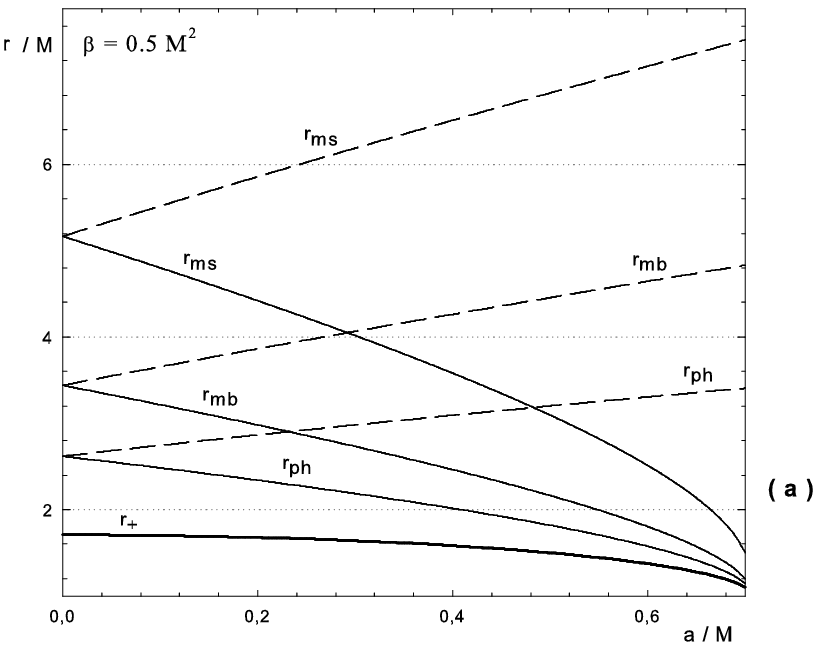}

~

~

\includegraphics[width=10cm,clip]{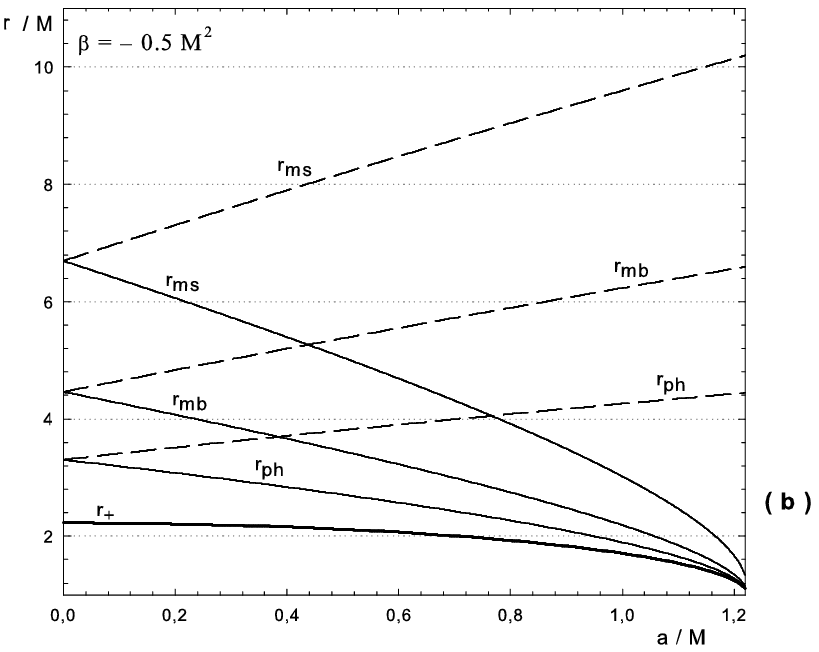}
\caption{ \label{geodesics,fixedbeta} The positions of the event
horizon, the limiting photon orbit , the marginally bound and the
marginally stable circular orbits around a rotating black hole on
the brane with the effects of the rotation parameter for given
values of $\,\beta =\pm\,0.5 M^2\,.$ The solid curves are for direct
motions, whereas the dashed curves are for retrograde motions.}
\end{figure}

\end{document}